\documentclass[aps,onecolumn,notitlepage]{revtex4-1}
\usepackage[utf8]{inputenc}
\usepackage[english]{babel}
\usepackage{amsmath}
\usepackage[dvipsnames]{xcolor}

\newcommand{\degg}[0]{^\circ}

\begin{document}

\title{Remarks on the use of objective probabilities in Bell-CHSH inequalities}

\author{Aldo F.G. Solis-Labastida}
\affiliation{Instituto de Ciencias Nucleares, Universidad Nacional Aut\'onoma de M\'exico, Apdo. Postal 70-543, C.P. 04510 CDMX, Mexico}
\author{Melina Gastelum} 
\affiliation{Facultad de Filosof\'\i a y Letras, Universidad Nacional Aut\'onoma de M\'exico, C.P. 04510 CDMX, Mexico}
\author{Jorge G. Hirsch}
\affiliation{Instituto de Ciencias Nucleares, Universidad Nacional Aut\'onoma de M\'exico, Apdo. Postal 70-543, C.P. 04510 CDMX, Mexico}

\begin{abstract}
The violation of Bell inequalities is often interpreted as showing that, if hidden variables exist, they must be contextual and non local. But they can also be explained questioning the probability space employed, or the validity of the Kolmogorov axioms. In this article we explore the additional constrains which can be deduced from two widely used objetive probability theories: frequentism and propensities. 

One of the strongest objections in the deduction of  one version of Bell inequalities goes about the probability space, which assumes the existence of values for the output of the experiment in each run, while only two of the four values can be measured each time, making them counterfactual. 
It is shown that frequentism rejects the possibility of using counterfactual situations, while long-run propensities allow their use. In this case the introduction of locality and contextuality does not help to explain the violation, and an alternative explanation could point to a failure of the probability.

Single case propensities were designed to associate probabilities to single events, but they need to be conditional to the whole universe, and do not have a clear link with the observed relative frequencies. It heavily limits their use.\\

\end{abstract} 

\maketitle

\section{Introduction}

From a minimalistic perspective, quantum physics can be described as a tool to assign probabilities to the outputs of measurements. While it is enough for its use in all kind of developments and technologies, it leaves open the question: what can be said about the processes described? Is it possible to model its properties?

The violation of Bell inequalities has been interpreted by different authors as giving some hints: if some unknown mechanism exist underlying this probabilistic behavior, it must be contextual and, if special relativity is valid, also non local.  
But we have recently stressed  that  {\em the hypothesis from which Bell inequalities are derived differ according to the probability space used to write them. The violation of Bell inequalities can, alternatively, be explained assuming that hidden variables do not exist at all, or that they exist but their values cannot be simultaneously assigned, or that the values can be assigned but joint probabilities cannot be properly defined, or that averages taken in different contexts cannot be combined} \cite{solis-labastida2021}.
We arrived to these conclusions formulating two versions of the Bell inequalities, which are formally equivalent, but are based on different probability spaces and on different hypothesis. 
There is no {\em a priori} criteria to select one of the possible formulations and, in view of the experimental observation of their violation, to decide which of the hypothesis must be rejected.

The previous analysis was, in line with many others, based in the axiomatic formulation of probability due to Kolmogorov \cite{gillies2000,rowbottom2015probability,kolmogorov2018foundations}. It provides solid mathematical basis, but leaves unanswered the question of to which situations a probability can be assigned, and how it can be interpreted.  

Different interpretations or extensions of quantum mechanics do have affinities for a particular, objective or subjective, probability theory \cite{Home1992}. 
Frequentism is invoked, in implicit or explicit forms, in ensamble interpretations, very often related with hidden variables theories \cite{Home1992} and in the application of Bell inequalities to certify random number generation \cite{fehr2013security,pironio2010random}, which in many cases assumes that the "creation" of values takes place in the measurement. Realist approaches which want to associate the quantum description with individual events, like those involving spontaneous collapse \cite{friggProbabilityGRWTheory2007}, are usually closer to an objective approach to probability like some form of propensities, while in the pilot-wave theory  \cite{callender2007} there is no explicit preference for a probability theory.

In the present contribution we select specific probability theories. Each of them helps to delimit the set of valid hypothesis needed to formulate the Bell inequalities. We analyze only objective probability theories, those who consider that probabilities are properties of the system, and not related with the information available or 
the beliefs of the agent assigning the probabilities. We include two paths to objective probability: the frequentist theory, widely used in the scientific communities, and different propensities. These theories provide useful elements for the analysis of Bell-CHSH inequalities.

The paper is organized as follows: after this introduction, the second section deals with the different possibilities for writing the inequalities. Section three discusses some characteristics of objective probability theories, and Section four analyses how the different inequalities fit in the different probability theories. Some conclusions are given at the end.

\section{Two Bell inequalities}

In our previous work \cite{solis-labastida2021}, we discussed two inequalities referred in the literature as Bell inequalities. They are formally equivalent but are based on different probability spaces and need different hypothesis to be proven. These inequalities are violated by experimental data, but the conclusions extracted from this violation depend on the description used.

We consider a classic Bell situation. A system is split up in two subsystems which are sent to different locations, say left and right. Such subsystems have two properties to be measured: $A,A'$ for the left side and $B,B'$ for the right side, each of which can only assume two values $0$ and $1$\footnote{
The tradition is to use values $+1$ and $-1$, however we use this values since its expectations values can be written directly as probabilities as can be seen in the following.
}. For each subsystem only one of their properties can be measured. Which one is measured is selected using an angle in each side: $\alpha=0\degg$ for $A$ and $\alpha=45\degg$ for $A'$, and the analogous for the right side.

The Bell-CHSH inequality in this situation is:
\begin{align}
    -1 \geq
    \langle AB \rangle
    + \langle AB' \rangle
    + \langle A'B \rangle
    - \langle A'B' \rangle
    - \langle A \rangle
    - \langle B \rangle
    \geq 0.
\end{align}
Since we consider only binary variables, each expected value $  \langle x \rangle$ can be written as the probability $P(x=1)$. They will be different probabilities depending on the probability space used. For our discussion we will consider two options.

The first inequality is based on a probability space containing four variables:
\begin{itemize}
    \item  $\alpha= 0\degg,45\degg$ The chosen polarization basis in the left hand side.
    \item $\beta= 0\degg,45\degg$ The chosen polarization basis in the right hand side. 
    \item $A= 0,1$ The result of the measurement in the left hand side.
    \item $B=0,1$ The result of a measurement in the right hand side.
\end{itemize}

This is probability space 1 or probability space $(A,\alpha,B,\beta)$. Such space leads to a Bell inequality with conditional probabilities:

\begin{align}
	-1 \leq \nonumber 
P(A=1,B=1|\alpha,\beta)  
&+P(A=1,B=1|\alpha',\beta) \nonumber \\
+P(A=1,B=1|\alpha,\beta') 
&-P(A=1,B=1|\alpha',\beta') \nonumber \\
-P(A=1|\alpha) 
&-P(B=1|\beta)  
\leq 0. 
\label{inequality1}
\end{align}

The hypothesis needed to prove such inequality are Kolmogorov axioms, $\lambda$-independence $P(\alpha,\beta|\lambda)=P(\alpha,\beta)$, and locality $P(A,B|\alpha,\beta)=P(A|\alpha)P(B|\beta)$ \cite{solis-labastida2021}.

The second inequality uses a different probability space. It consists of:
\begin{itemize}
    \item $A_{0\degg}$ The result in  the left hand side if the measurement would be set to $0\degg$.
    \item $A_{45\degg}$ The result  the left hand side if the measurement would be set to $45\degg$.
    \item $B_{0\degg}$ The result in the right hand side if the measurement would be set to $0\degg$.
    \item $B_{45\degg}$ The result in the right hand side if the measurement would be set to $45\degg$.
\end{itemize}

This is probability space 2 or probability space $(A_{0\degg},A_{45\degg},B_{0\degg},B_{45\degg})$. In this case a different Bell-CHSH inequality is obtained:

\begin{align}
	-1 \leq 
	P(A_{\alpha}=1,B_{\beta}=1) 
	&+P(A_{\alpha}=1,B_{\beta'}=1) \nonumber \\
	+P(A_{\alpha'}=1,B_{\beta}=1) 
	&-P(A_{\alpha'}=1,B_{\beta'}=1) \nonumber \\
	-P(A_{\alpha}=1) 
	&-P(B_{\beta}=1) 
	\leq 0.
\label{inequality2}
\end{align}

This inequality is only based in Kolmogorov axioms. In its deduction there is no need to invoke locality or $\lambda$-independence.

Both probability spaces properly describe the experimental situation when probability is understood only in terms of Kolmogorov axioms. The observed violation of these inequalities implies that at least one of the hypothesis in which they are based must be refuted. Employing the probability space 1 and the validity of the Kolmogorov axioms lead many authors
to conclude that any hidden variable model must be non-local and/or contextual. On the other hand, probability space 2 only offers interpretations regarding the use of probability: that values cannot be simultaneously assigned to all variables, or that the values can be assigned but joint probabilities cannot be properly defined, or that such situations pertain to different probability spaces \cite{solis-labastida2021}.

The purpose of this paper is to analyze the consequences of the violation of Bell inequalities in the context of two widely used (explicit or implicitly) objective probability theories:  frequentism and propensities. In the next section we review some of their properties.

\section{Objective probabilities}

Historically there are two main approaches to probability: \emph{epistemic} and \emph{objective} theories. The epistemic approach has a fundamental role in probability and was the most used in the XIX century. The concepts of probability employed today in the scientific community, from education to practice, are strongly rooted in these theories, although this fact is not always recognized. It considers probability as fundamentally related with the knowledge (or the beliefs) a person has about the events whose probabilities are assigned. Examples are the classical theory of probability, the logicist theory of probability, and the subjective theory of probability\cite{gillies2000}. 
Objective theories, on the other hand, consider probabilities an objective property, like length or mass. Two relevant members of this group are the frequentist theory of probability and propensities. 

In what follows some important characteristics of these two branches of probability are presented, in order to relate them later to Bell-CHSH inequalities.

\subsection{ Frequentism }

Frequentism is a theory of probability that has different versions, being the most famous those formulated by Richard von Mises and by Hans Reichenbach. In this work we will consider frequentism in the view of von Mises \cite{mises1981}.

Frequencies, from where the term frequentism derives, refer to the number of times a situation repeats. The classical picture is the description of an experiment, game, or other situation, where different outcomes are possible. 
After many repetitions, each particular outcome has appeared a number of times. This number is called its frequency, and is divided by the total number of repetitions. This ratio is called the {\em relative frequency}.

These repetitions are always characterized by a set of parameters used to identify the events. In frequentism this set is called the \emph{attribute space}, similar to the sample space in the Kolmogorov formalism  \cite{gillies2000}. In a coin the attributes are "heads" and "tails", in a dice the number of dots in the face upwards. 
 
The fundamental fact for frequentism is that relative frequencies acquire a stable value as the number of repetitions increase. It is important to note that not all the situations have this property. However, many common situations,  particularly situations of interest to scientists,  do have this property. Experiments confirming the stability of relative frequencies for coins and other games have been performed along the history of probability \cite[p.152]{gillies2000}.

The key element for such convergence is a huge number of repetitions, which can be performed employing a large number of coins, or the same coin over and over again. This is the first and most important element of frequentism. There must be a sequence called a \emph{collective}, \begin{quotation}
    ...it denotes a sequence of uniform events or processes which differ by certain observable attributes, say colours, numbers, or anything else.\cite[p.12]{mises1981}
\end{quotation} Although in practice it is only possible to have a finite number of elements in an actual sequence, frequentism postulates that collectives have an infinite number of elements, and that the relative frequencies will converge to a fixed value. Such situation is captured using the axiom of convergence \cite{mises1981}.  If $N(A)$ is the number of times an attribute $A$ appears in a collective, the axiom states that, for any collective, the relative frequencies have a limit as the number of elements tends to infinity, i.e. the limit
\begin{align}
    P(A) \equiv \lim_{n \to \infty} \frac{N(A)}{n}
\end{align}
exists and is by definition the probability $P(A)$ of the occurrence of the event $A$.

Within this approach the probability of a single event has no meaning, because the probability is a property of a collective \cite{gillies2000,mises1981}. There is no place to sentences like ''the probability that this afternoon will rain", as this event will take place, or not, only once: today's afternoon. On the other hand, it is perfectly appropriate to define the probability of raining in October, or the probability of rain on Tuesdays in October, where a collective can be built considering all the Tuesdays in October in the last 100 years. But notice that the association of a particular event to a collective can be performed in many different ways, and quite possibly a different probability will be assigned  depending on the collective selected. 

The other important element in frequentism is the \emph{randomness condition} \cite{mises1981}. Given a converging sequence, it is possible to arbitrarily select some of its members, say all even positions. It is expected that both sequences have the same limiting value of their relative frequencies. There should not be a method to build subsequences that changes the limit value. The mathematical work required  to properly describe the randomness condition is out of the scope of this work, for a comprehensive account see \cite{vanlambalgen1996randomness}. It is worth to mention that in many common situations it is not possible to fulfill this condition. 

\subsection{Propensity approach to probability}

The propensity approach to probability was intensively worked along the second half of the XX century. It is due to Popper \cite{popper1959,popper1967}, who wanted to build an objective account of probability to deal specifically with quantum mechanics. Since then, different versions have been developed.

Frequentism offers an objective view of probability, but it explicitly denies the use of probability in individual phenomena. This is what propensities tried to introduce in its inception \cite{popper1959,popper1967}. Nowadays, as Gillies points out\cite{gillies2000}, the term is used somewhat freely to describe any objective theory not based in frequencies.

The original idea is to attribute probability, not to collectives, but to the conditions on which a collective is produced. Therefore, probabilities can be seen as a measurement of the tendency, or propensity (hence the name), those conditions have to produce a certain outcome. In this way the probability of a single event is identified with the probability of those conditions on which the event takes place. This is explicitly said by Popper:
\begin{quotation}
    ...probability may now be said to be \emph{a property of the generating conditions }[of a sequence].\cite{popper1959}
\end{quotation}
A more explicit passage says the same a few years later:
\begin{quotation}
    \emph{Propensities are properties of neither particles nor photons nor electrons nor pennies}. They are \emph{properties of the repeatable experimental arrangement}...\cite{popper1967}
\end{quotation}

This is important since all probabilities will be, in a sense, conditional probabilities. All are conditioned by the situation in which the event takes place. 

This characteristic is what Popper uses to assign probabilities to single events, the probability of a single event is that of its  experimental conditions.
\begin{quotation}
    For now we can say that the singular event $a$ possesses a probability $p(a, b)$ owing to the fact that it is an event produced, or selected, in accordance with the generating conditions $b$.\cite{popper1959}
\end{quotation}

The problem with this proposal is that a single event can be characterized under different sets of conditions. An alternative option is to let go probabilities of individual events but still hold that probabilities are properties of conditions. Both paths have been explored in the last fifty years.
The classification of propensity theories is due to Gillies. Propensities with objective probabilities for individual events are called \emph{single case theories} while the others are called \emph{long run theories}\cite{gillies2000}. 

The goal of single case theories is to assign objective probabilities to single events. These is the path Popper chose~\cite{popper1990world}. 
To fully characterize a single event all the conditions in the world at that moment should be specified. But in this case this event is unique, there is no other like it, because similar events will take place in different locations or in different moments. For this reason, it is impossible to determine the relative frequency of this kind of individual events. If the probability of this event "exists objectively", there is no way to determine it.

On the other hand long-run theories will accept probability is attributed to conditions. However, propensities will measure the propensity those conditions have to produce relative frequencies when repeated many times. Gillies and others took these approach\cite{gillies2000,hacking2016}.

For instance, if a coin is tossed, the probability to obtain heads or tails is:
\begin{align}
P(heads| S), \nonumber \\
P(tails| S)
\label{proba1}
\end{align}
where $S$ represents the conditions. In a single-case theory, $S$ must include all the conditions of whole world for a particular toss. If the coin can be catched with the right or the left hand, this condition must be explicitly added: 
\begin{align}
    P(heads|right \land S) \nonumber \\
    P(tails|left \land S),
\end{align}
since a single case theory must have exhaustive conditions.

In the case of a long run theory, the set of conditions may not be exhaustive. 
A long-run theory can have Eq. (\ref{proba1}) without mentioning which hand is used to catch the coin, since conditions do not have to be exhaustive.

\section{The Bell-CHSH inequalities and different probabilities}

The conclusions extracted from the observed violation of the Bell inequalities depend on the probability theory employed. 
In some cases, the probability space 2 is not acceptable, and the only valid inequality is the one built in space 1. 

\subsection{Frequentism}

In frequentism the attribute space must classify all the possible outcomes of an experiment. When talking about games this seems an easy task, because the (explicit or implicit) rules of the game define the attributes as the number of dots in the dice, the face in the coin, etc. Any other output will be excluded from the valid results.
But in most situations, in life and in science, there are many options, and it is possible to classify events according to different criteria. 

Referring to the Bell-CHSH experiment, the  sample space 1, $(A,\alpha,B,\beta)$, allows to characterize  the different outcomes of the experiment. That is, every experiment has a unique quadruple that characterize it.  For sample space 2, $(A_{0\degg},A_{45\degg},B_{0\degg},B_{45\degg})$, the situation is different. First, an experimental outcome cannot be classified as one of the elements of the sample space. If in a run  $A_{0\degg}$ and $B_{0\degg}$ are measured, $A_{45\degg}$ and $B_{45\degg}$ are unknown. 

The strongest objection to the use of sample space 2 from the frequentist view is counterfactuality: in a Bell-CHSH scenario it is accepted that $A_{0\degg}$ and $A_{45\degg}$ cannot be measured simultaneously, these events are  mutually exclusive. 
This is not exclusive to quantum mechanics. There are common situation where only one of two options can be taken. Von Mises refers to such situations in his text:
\begin{quotation}
    Consider a good tennis player. He may have 80\% probability of winning in a certain tournament in London. His chance of winning another tournament in New York, beginning on the same day, may be 70\%. The possibility of playing in both tournaments is ruled out, hence, the events are mutually exclusive, but it is obviously nonsense to say that the probability of his winning either in London or in New York is 0.80 + 0.70 = 1.50. In this case again, the explanation of the paradox lies in the fact that the two probabilities refer to two different collectives, whereas the addition of probabilities is only allowed within a single collective.\cite{mises1981}
\end{quotation}

If in a Bell-CHSH situation $A_{0\degg}$ and $A_{45\degg}$ cannot be measured simultaneously, so the two measurements must be considered as mutually exclusive, it can be argued they have to be considered as a different collectives, just like the previous example. 
Therefore the probability $P(A_{0\degg},B_{0\degg})$ and the probability $P(A_{45\degg},B_{0\degg})$ in equation \ref{inequality2} correspond to different collectives and should not be added. This represents a problem for inequality 2, eq.\ref{inequality2}, since each term with two variables corresponds to different collectives.

The previous argument then invalidates inequality 2 as a correct use of frequentist probability or, from another view, can explain the violation via a misuse of probability. The application, explicit or implicit, of the frequentist theory to Bell inequalities can be found in \cite{cetto2020spin,shimony1984contextual,svetlichny1988,vervoort2011,nieuwenhuizen2011contextuality} and others. 

\subsection{Propensities}

The analysis of the violation of Bell inequalities using propensities must be performed from a different perspective. As mentioned above, all propensities are conditional to the experimental situation. 

\subsubsection{Single case propensities}

Single case propensities are used to assign objective probabilities to individual events.

In equation \ref{inequality2}, $P(A_{0\degg},B_{0\degg})$ and $P(A_{45\degg},B_{45\degg})$, from this perspective, refer to the same pair of particles, they are not the probability of a group. The problem in this case is that there is no clear connection with the experiment. Each individual measurement has different conditions, at least because they take place at different times, and the formalism does not allow for the interpretation of the probabilities with the relative frequencies observed. 

A specific objection to the use of probability space 2 is that single case propensities need to be conditioned to the whole experimental situation. 
The terms in inequality 2 have a different conditions on the experimental apparatus and therefore must have different conditionals:
\begin{align}
    P(A_{0\degg},B_{0\degg} | \alpha=0\degg \land \beta=0\degg \land S), \nonumber \\
    \vdots \nonumber \\
    P(A_{45\degg},B_{45\degg} | \alpha=45\degg \land \beta=45 \land S'),
\end{align}
where $S, S'$ represent in each  case all other conditions in the world. The polarization angle must be part of the conditional. At first view, it seems to imply that, using single case propensities, probability space 1 is more appropriate, or, alternatively, the angle variables must be included as conditionals to the space 2, employing the  probability space 3 defined in Ref.  \cite{solis-labastida2021}. 

But there is an alternative option. Miller, in his propensity account explicitly says to which conditions a propensity should be conditioned:
\begin{quotation}
    Strictly, every propensity (absolute or conditional) must be referred to the complete situation of the universe (or the light-cone) at the time.\cite{miller2015critical}
\end{quotation}

As a plethora of different experimental variations of the Bell experiment have been performed where the selection of the polarization basis and the detection on one side is outside the light cone of the detector on the other side
(\cite{barrett2002quantum,gisin1999bell,hensen2015loopholefree} and many others), it would allow for the use of the same conditionals for all probabilities, and the sample space 2 can be reconciled with single case propensities.

\subsubsection{Long run propensities}

Long run propensities are conditioned to the experimental situation, but such conditions do not need to be exhaustive, i.e. it is not necessary to have a complete specification of the world. The propensity :
\begin{align}
    P(A_{0\degg},B_{0\degg}| S)
\end{align}
has the conditions $S$, but such conditions are not forced to contain the disposition of the experimental apparatuses.
Hence the conditional can be the same for all the probabilities and inequality 2 can be used.

\subsection{A broader view}

To conclude this section, is important to note that all the above arguments deal with probability space 2. Probability space 1 appears to have no issues, which is expected since there is no counterfactual situation in this space. That is, it is possible to assign values to all the variables in the space using the result of the experiment in a single experimental run. Employing these arguments the focus goes to inequality 1, where it is possible to maintain the probability space and concentrate in locality or $\lambda$-independence as the main reasons why a violation of inequalities is observed.

While the use of probability space 2 is severely questioned in the frequentist view, and has difficulties with propensities,   it is widely employed in the quantum information community, for instance in random number generation \cite{pironio2010random}.The guaranteed secrecy and randomness in these procedures is based on assuming that the violation of inequality 2 implies the non-existence of hidden variables. But it is worth to mention that these hidden variables could exist, but do not have a joint probability, or the other options discussed in \cite{solis-labastida2021}.

\section{The probability space and the Kochen-Specker theorem}

Bell inequalities relate a probabilistic analysis with experimental results. There is no mention to quantum physics. It gives them a special place in physics, because their violation puts limits which any theory aimed to describe these experiments should fulfill.   
As quantum theory has been employed successfully to describe a wide variety of observations and technological developments, it is tempting to assume its validity, and how it limits the universe of possible hidden variables theories. 

The Kochen Specker (KS) theorem is relevant  in this context. It requires that observables fulfill the algebra of quantum mechanics, which is a very reasonable hypothesis, but it means that we accept the experiment is well described by quantum mechanics, adding another hypothesis to our list. 

KS theorem states that a non-contextual assignment of values that meets the requirements of the quantum mechanics is impossible. It relies in two hypothesis regularly known as non-contextuality and values. Non-contextuality states that the assignment of values to observables does not depends on which set of observables is measured. Values states that it is possible to assign a value to every observable. Certainly, this section only presents a brief account of KS theorem, we invite the reader to consult \cite{cabello2021} to get a broader view on the subject.

Since there have been arguments against KS theorem \cite{barrett2004,meyer1999}, including a mathematical construction capable to assign non-contextual values \cite{pitowsky1989}, it is necessary to introduce probabilities in the analysis of the KS theorem. A different characterization of contextuality has been presented in \cite{appleby2002,cabello2002}, using probability as their framework and have produced sets of inequalities, closely related to inequality 2, whose violations characterize the contextual behavior of a system \cite{cabello2002,araujo2013all}. However we have to highlight that this is mathematically independent of the KS contextuality as is clearly stated in \cite{cabello2002}.

As pointed in \cite{budroni2021} such discussion has been very important since 
\begin{quote}
    These works played a fundamental role in the development of the modern approach to contextuality by stimulating the extension of the KS notion of contextuality from a logical to a probabilistic framework.
\end{quote}

In conclusion, a more robust characterization of contextuality is based in probability. To use such description in this analysis will require to use the same theory of probability in order to be coherent in our approach. Since this exceeds the scope of this work we restrict ourselves to the analysis of the  KS theorem.

At this point, the objections raised on the use of probability space 2 can be formulated in the terms of the KS theorem.
For inequality 2 the value of two different directions, for each particle, is needed. Therefore, if values is not satisfied, some of the directions used may not have an assigned value making the description of probability space 2 incorrect. 

If values holds but non-contextuality not, i.e. the values of observables depend on which set of observables is measured, probability space 2 can be employed. In this case the values used are granted, but there is no element in the probability space that allows to determine the variables which were not measured. This is not a problem for axiomatic probability, long-run propensity, and frequentism. In all these approaches the probability space does not has to have all the variables involved and only needs to characterize the outcomes. Returning to the example of a coin, the sample space of a coin only uses \emph{heads} and \emph{tails}, although it is accepted that the initial positions and velocities are important variables which determine the final result. 
  
\section{Conclusions}

When interpreting the violation of Bell-CHSH inequalities, probability plays a fundamental role. There are different probability spaces which can be employed to deduce the inequalities, with different hypothesis involved. The observed violation forces to reject at least one of these hypothesis, opening the space for different possible interpretations. Some communities conclude that they exhibit that, if hidden variables exist, they must be contextual and non-local. For others it guarantees that the results of the measurements are created at this moment, and are absolutely random. But there are other options: it could be that the hidden variables exists, but it is not possible to assign a joint probability to them, or that one of the Kolmogorov axioms is not valid \cite{solis-labastida2020}. 

In this article we explored the different constrains to the probability spaces which can be deduced from two widely used objetive probability theories: frequentism and propensities. The strongest objections go over the probability space 2, which assumes the existence of values for the output of the experiment in each run, while only two of the four values can be measured each time, making them counterfactual. 

Frequentism, in von Mises version, rejects the possibility of using counterfactual situations so inequality 2 will not be a proper use of this probability theory. This only leaves inequality 1 for a proper analysis of a Bell-type experiment, and locality and $\lambda$-independence can be used to explain the experimental violations observed. 

On the other hand long-run propensities, those with fewer problems in their relation to relative frequencies, will allow the use of both inequalities, 1 and 2. In order to explain their violation, the introduction of locality and $\lambda$-independence only helps with inequality 1. For inequality 2  an explanation related with a failure of the probability spaces, the non existence of a joint probability, is needed.
It is worth mention that if both inequalities are accepted as valid descriptions of the experiment, and we want to use only one reason to explain both, it has to be related with probability. 

Single case propensities were designed to associate probabilities to single events, but they need to be conditional to the whole universe, and do not have a clear link with the observed relative frequencies. It heavily limit their use.   

The diversity in arguments when interpreting the inequalities \cite{solis-labastida2020} match the communication barrier between communities holding different interpretations. That can be attributed to the different inequalities considered as Bell-CHSH inequalities and the different probability theories used, or the lack of their explicit identification.

Declarations of interest: none.

Funding: This research was partially funded by DGAPA-UNAM project IN104020.

CRediT author statement: Aldo Solis: Investigation, conceptualization, Writing- Original draft preparation, Jorge Hirsch: Conceptualization, writing, reviewing, editing. Melina Gastelum:  Conceptualization, writing, reviewing, editing.

\bibliographystyle{unsrt}

\end{document}